# DeGaPe 35: Amateur discovery of a new southern symbiotic star

Thomas **Petit**[a,b], Jaroslav **Merc**[c,d,*], Rudolf **Gális**[d], Stéphane **Charbonnel**[a], Thierry **Demange**[b], Richard **Galli**[b], Olivier **Garde**[a], Pascal **Le Dû**[a] and Lionel **Mulato**[a]

[a]*Southern Spectroscopic Project Observatory Team (2SPOT), France*
[b]*Atacama Photographic Observatory (APO), France*
[c]*Astronomical Institute, Faculty of Mathematics and Physics, Charles University, V Holešovičkách 2, Prague, 180 00, Czech Republic*
[d]*Institute of Physics, Faculty of Science, P. J. Šafárik University, Park Angelinum 9, Košice, 040 01, Slovak Republic*

ARTICLE INFO

ABSTRACT

In this work, we present the discovery and characterization of new southern S-type symbiotic star, DeGaPe 35. We have obtained the low-resolution spectroscopic observations, and supplemented it by the photometry from *Gaia* DR3 and other surveys. The optical spectra of this target show prominent emission lines, including highly ionized [Fe VII] and O VI lines. The cool component of this symbiotic binary is an M4-5 giant with $T_{\text{eff}} \sim 3\,380 - 3\,470\,\text{K}$ and luminosity $\sim 3 \times 10^3\,\text{L}_\odot$ (for the adopted distance of 3 kpc), the hot component is a shell-burning white dwarf. The photometric observations of the *Gaia* satellite, published recently in the *Gaia* DR3 suggested the variability with the period of about 700 – 800 days, that we tentatively attributed to the orbital motion of the binary.

## 1. Introduction

The growing availability of affordable spectrographs for amateur observers in recent years increases the opportunities for amateurs to participate in the scientific research of many interesting objects in the night sky. One of the fields that benefits significantly from the long-term, high-cadence observations of amateur astronomers is the research of symbiotic binaries (see, e.g., Iłkiewicz et al. 2016, 2022; Skopal et al. 2017, 2020; Lucy et al. 2020; Shagatova et al. 2021; Merc et al. 2021b, 2022; Pandey et al. 2022, and the discussion in Merc et al. 2021a).

These interacting systems consist of a cool giant of spectral type M (or K) that is loosing its mass via the stellar wind or the Roche-lobe overflow and a hot component accreting the matter. In most cases, it is a white dwarf surrounded by extensive pseudo-photosphere with high quiescent temperature ($\sim 10^5$ K) and luminosity ($\sim 10^{2-4}\,\text{L}_\odot$) due to stable thermonuclear burning of accreted hydrogen-rich matter. The optical spectra of such objects have very prominent emission lines that allows them to be detected even in the narrow-band photometric surveys (see, e.g., reviews by Mikołajewska, 2012; Munari, 2019). There are also other objects, which share some of the photometric and/or spectroscopic properties with symbiotic stars, e.g., the strong emission lines. Often, objects previously though to be planetary nebulae are reclassified as symbiotic stars based on the more detailed follow-up observations (see, e.g., Allen, 1984; Acker et al., 1987; Frew and Parker, 2010).

DeGaPe 35 (= 2MASS J15211785-5900339; $\alpha_{2000}$ = 15:21:17.86, $\delta_{2000}$ = -59:00:33.90; see also Tab. 1), previously anonymous field star, was identified as a possible emission-line object in the scope of an amateur survey searching for planetary nebulae (Le Dû et al., 2018, 2022). This survey is operated as a supplementary program at the amateur-built, remotely-operated Atacama Photographic Observatory (APO)[1] located in the San Pedro de Atacama, Chile. The main APO instrument is a Takahashi TOA 15-cm refractor telescope, which is equipped with the *LRGB* broadband filters, narrow-band H$\alpha$, S II, and O III filters and the Apogée Alta U16M cooled camera. The main goal of the observatory is the astrophotography of the southern deep-sky objects. However, the field of view of four square degrees makes the obtained data excellent to search for conspicuous objects in the vicinity of the photographed targets.

*Corresponding author
✉ petittom@gmail.com (T. Petit); jaroslav.merc@gmail.com (J. Merc); rudolf.galis@upjs.sk (R. Gális)
ORCID(s): 0000-0001-6355-2468 (J. Merc); 0000-0003-4299-6419 (R. Gális)

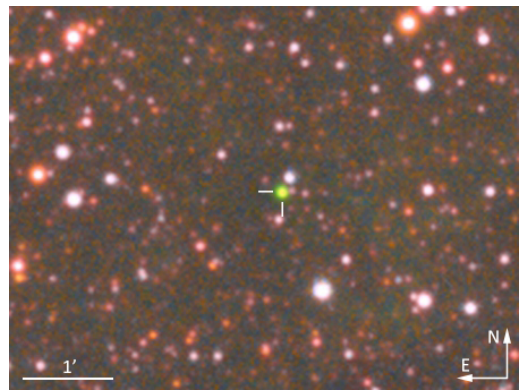

**Figure 1:** Discovery image of DeGaPe 35. The target is the conspicuous green object in the middle. See the text for more details.

DeGaPe 35 was discovered in the field of the well-known Wolf-Rayet star WR 68 during our observations on May 9, 2017. The final discovery image is the combination of H$\alpha$, S II, and O III observations with total exposure times of 800, 600, and 600 minutes, respectively. DeGaPe 35 was identified thanks to its green appearance on the image (Fig.

---
[1]http://www.atacama-photographic-observatory.com/index_EN.php

**Table 1**
Basic properties of DeGaPe 35. Data are from Gaia DR3 (Gaia Collaboration et al., 2022), 2MASS (Skrutskie et al., 2006), and WISE (Wright et al., 2010).

| Parameter | Value |
|---|---|
| $\alpha_{2000}$ [h:m:s] | 15:21:17.86 |
| $\delta_{2000}$ [d:m:s] | -59:00:33.90 |
| $l_{2000}$ [deg] | 321.191 |
| $b_{2000}$ [deg] | -1.557 |
| $\mu_\alpha$ [mas/yr] | -4.347 ± 0.061 |
| $\mu_\delta$ [mas/yr] | -2.839 ± 0.062 |
| $\pi$ [mas] | 0.197 ± 0.053 |
| $G$ [mag] | 13.303 ± 0.004 |
| $BP$ [mag] | 16.212 ± 0.032 |
| $RP$ [mag] | 11.772 ± 0.007 |
| $J$ [mag] | 8.339 ± 0.026 |
| $H$ [mag] | 6.877 ± 0.027 |
| $K_S$ [mag] | 6.248 ± 0.026 |
| $W1$ [mag] | 5.921 ± 0.044 |
| $W2$ [mag] | 5.920 ± 0.020 |
| $W3$ [mag] | 5.605 ± 0.020 |
| $W4$ [mag] | 4.889 ± 0.045 |

1). Such an appearance is typical for stellar-sized planetary nebulae and other emission-line objects which have strong emission lines in their optical spectra, especially [O III] and H$\alpha$ lines. Objects with such an appearance are identified on the images and included to the HASH database of planetary nebulae (Parker et al., 2016) and to the French amateur planetary nebulae catalog[2] as possible new planetary nebulae. The candidates are subjected to spectroscopic follow-up to confirm the planetary nebula nature or reclassify them as mimics. In the case of DeGaPe 35, our follow-up observations in June 2021 (see below) contradicted the planetary nebula nature of the source and the object was reclassified as an emission-line star in the afore-mentioned databases. We reviewed these observations in the scope of the study on the symbiotic candidates (see Merc et al., 2021b) and claim, that the object is not only the emission-line star, but can be classified as a symbiotic star. The paper is structured as follows: in Section 2, we describe our and archival observational data used in this work, in Section 3, we investigate the symbiotic nature of the studied source, including its symbiotic-like variability and in Section 4, we infer the parameters of the binary components.

## 2. Observational data

We have obtained two low-resolution spectra of DeGaPe 35 using the Southern Spectroscopic Project Observatory[3] remotely-operated Ritchey-Chrétien 12" telescope located at the Deep Sky Chile facilities, equipped with an Alpy600 spectrograph (23 $\mu$m slit, providing a resolving power ∼ 550) and Atik 414EX cooled camera. The first spectroscopic data were obtained on June 11, 2021 (JD 2 459 376.6), the second dataset was acquired on May 10, 2022 (JD 2 459 709.8). In both cases, total of 6 exposures of 1 200 s was summed to produce the final spectra (see Fig. 2). The spectra of argon-neon calibration lamp were acquired right after the target for the wavelength calibration (typical FWHM of argon-neon lines ∼ 10 – 12 Å). For the evaluation of the instrumental response (together with the atmospheric effects), the spectra of reference star (HD 142139), located on the sky close to DeGaPe 35, were obtained afterwards (18 exposures of 10 second each in both cases), followed by another set of spectra of the argon-neon calibration lamp. Acquisition of the data was done with Prism V10.4.12.911. The spectra were reduced with ISIS V6.1.1 software. Fully processed spectra are produced after dark, bias, flat-field, instrumental correction, wavelength calibration, and atmospheric lines and cosmic rays removal (see also Le Dû et al., 2022).

For the analysis of SED of DeGaPe 35, we have collected the data from the *Gaia* DR3 ($G_{BP}$, $G$, $G_{RP}$; Gaia Collaboration et al., 2022), SkyMapper Southern Survey (*g*, *r*; Wolf et al., 2018), Two Micron All-Sky Survey (2MASS; *J, H, K*; Skrutskie et al., 2006), Midcourse Space Experiment (*MSX*; *A* band; Price et al., 2001), and *Wide-field Infrared Survey Explorer* (*WISE*; *W1, W2, W3, W4*; Wright et al., 2010).

We have also searched for the time-resolved photometric observation of the target to analyze the possible variability of the target. Unfortunately, the photometry of DeGaPe 35 from the ASAS-SN survey (Shappee et al., 2014; Kochanek et al., 2017) is contaminated by the nearby bright sources, and the target is located on the southern part of the sky not covered by the ZTF survey (Masci et al., 2019). The brightness of the star seems to be close or below the sensitivity limit of the ASAS survey (Pojmanski, 1997). The only usable photometric measurements are those obtained by the *Gaia* satellite in the $G_{BP}$, $G$, $G_{RP}$ filter and published in the *Gaia* DR3 (Gaia Collaboration et al., 2022). However, these cover only the interval of 34 months (July 2014 – May 2017), which might not be sufficient to study the orbital variability of symbiotic stars (see, e.g., Gromadzki et al., 2013).

## 3. Symbiotic nature

The studied object was classified as an emission-line star (without any further specification) in the HASH database and the French amateur planetary nebulae catalog. However, already at the time when the object was included in the later catalog, the possible presence of the Raman-scattered O VI lines was noted in the optical spectra. Our analysis of the low-resolution optical spectra of DeGaPe 35 (Fig. 2) confirmed the presence of the Raman-scattered O VI emission lines, which is generally a sufficient condition for a symbiotic classification in conjunction with the presence of a cool giant in the system (Belczyński et al. 2000; Akras et al. 2019a; see also Nussbaumer et al. 1989; Schmid 1989). Moreover, other highly ionized emission lines have been detected in the spectrum, including the prominent emission lines of He II, and [Fe VII], in addition to strong emission

---

[2]http://planetarynebulae.net/EN/index.php
[3]https://2spot.org/EN/

DeGaPe 35: New symbiotic star

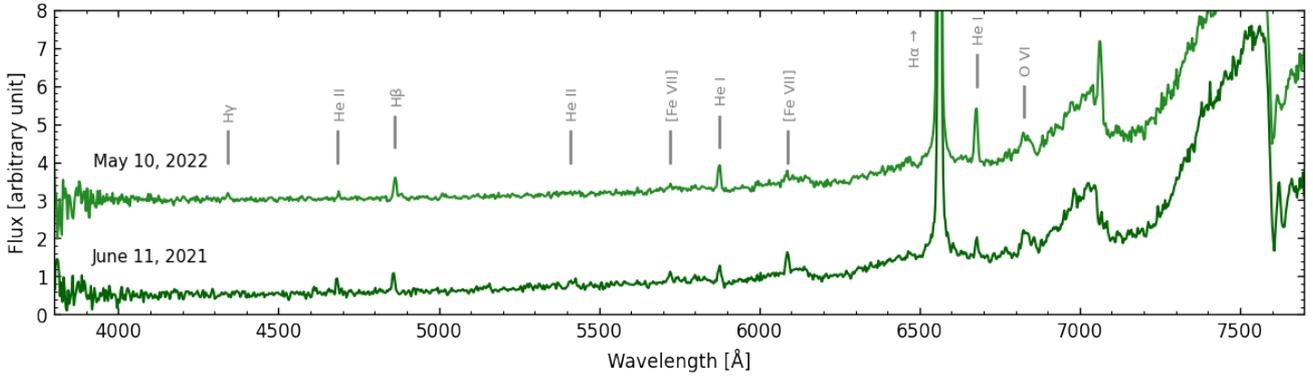

**Figure 2:** Low-resolution spectra of DeGaPe 35. The spectra obtained on June 11, 2021 and May 10, 2022 are depicted by the dark and light green color, respectively. Spectra are separated by the arbitrary constant. The identification of prominent emission lines detected in the spectra is given by the vertical lines.

lines of H I and He I. The continuum can be well-fitted by the spectrum of a cool M-type giant (M5 III, see below). Taken together, the optical spectra of DeGaPe 35 satisfy nowadays most often accepted symbiotic criteria of Belczyński et al. (2000), which are the presence of the absorption features of a late-type giant in the spectrum and of the emission lines of ions with an ionization potential (IP) of at least 35 eV. We should note that the obtained spectra satisfy also the more stringent definition first introduced by Allen (1984).

The symbiotic classification is further supported by the IR colors of DeGaPe 35 that satisfy the IR criteria for S-type symbiotic binaries (Akras et al., 2019b, 2021). Moreover, the object is classified as a symbiotic binary in all seven classification trees suggested by Akras et al. (2019b) to distinguish between symbiotic stars and common mimics. The lack of [O III] lines in the spectra of DeGaPe 35 precludes the use of diagnostic diagrams based on the [O III] and Balmer lines fluxes (e.g., Iłkiewicz and Mikołajewska, 2017; Iłkiewicz et al., 2018). We have employed the He I diagram (Fig. 4 in Iłkiewicz and Mikołajewska, 2017) in which DeGaPe 35 is located in the region occupied solely by symbiotic stars (for the spectrum obtained in May 2022: log(He I $\lambda 6678$/He I $\lambda 5876$) = -0.10, log(He I $\lambda 7065$/He I $\lambda 5876$) = -0.12; spectrum from 2021 was not used as the He I $\lambda 7065$ was faint).

### 3.1. Variability

The unavailability of the long-term photometric data doesn't allow the detailed analysis of the variability of DeGaPe 35. The light curves from *Gaia* DR3 (Gaia Collaboration et al., 2022), shown in Fig. 3, suggest the variability on the timescale of 700 – 800 days, but the covered time interval is too short for a proper period analysis. The amplitude of these changes is larger in $G_{BP}$ compared to the $G_{RP}$ filter, and is about 1 and 0.2 mag, respectively. The timescale and the color dependence of the changes are similar to the orbital variability observed in other S-type symbiotic binaries (e.g., Gromadzki et al., 2013). Therefore we tentatively attribute these brightness changes to the orbital motion of the symbiotic binary.

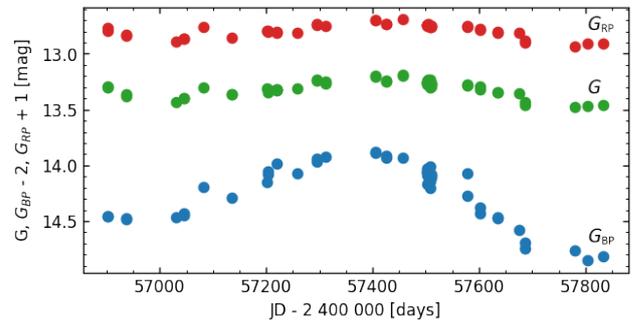

**Figure 3:** *Gaia* DR3 light curves of DeGaPe 35. The observations in $G_{BP}$, $G_{RP}$, and $G$ filters are shown in blue, green, and red, respectively. For clarity, the data in $G_{BP}$ and $G_{RP}$ were shifted by constants of -2 and +1 mag, respectively.

Also the spectroscopic data obtained eleven month apart show some changes in the strengths of the emission lines (Fig. 2). Namely, the intensity of H I and He I lines increased while the He II and [Fe VII] lines are weaker in the later spectrum. Such a changes are often detected in the symbiotic binaries and are generally caused by the orbital motion or the changes in the ionization structure of the nebula.

## 4. Components of the binary
### 4.1. Cool component and the distance

In the optical spectra (Fig. 2), the contribution of the red giant is well visible thanks to the prominent molecular bands of the TiO. We have employed the method of Kenyon and Fernandez-Castro (1987, equations 1 and 2) to estimate the spectral type of the cool component using the strengths of TiO bands. For the two low-resolution spectra, we have obtained [TiO]$_1$ = 0.3 – 0.4 and [TiO]$_2$ = 0.6 – 0.8. This corresponds to the spectral type $\sim$M2-4 according to the original calibration of Kenyon and Fernandez-Castro (1987, their figures 2 and 3). However, the resolution of the spectra used by the authors is higher than of the spectra we have obtained for DeGaPe 35. Therefore, we have calibrated

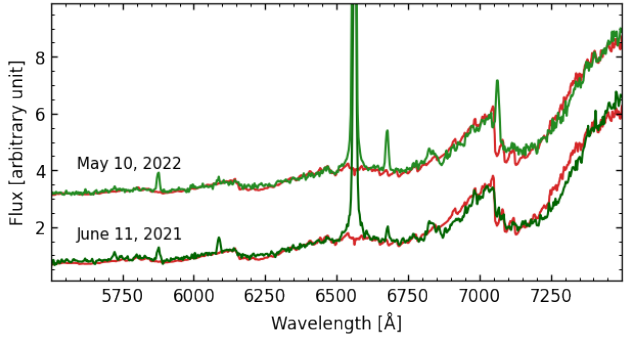
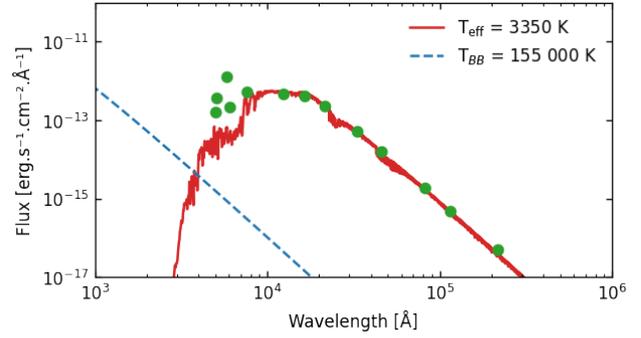

**Figure 4:** Comparison of observed low-resolution spectra of DeGaPe 35 (shown in green) with the intrinsic spectrum of M5 III star (shown in red; Fluks et al., 1994). The reference spectrum was reddened by the value of $E(B-V) = 1.7$ mag (see the text for details).

**Figure 5:** Multi-frequency SED of DeGaPe 35 constructed on the basis of data from the *Gaia* EDR3, SkyMapper, 2MASS, MSX, and WISE (in green). The best-fitting theoretical spectrum with $T_{\rm eff} = 3\,350$ K and $\log g = 0.0$ is shown in red (Allard, 2014). The dashed blue line denotes the radiation of a blackbody in a distance of 3 kpc with a temperature of $1.55 \times 10^5$ K, and luminosity of $2\,350\,L_\odot$ (a hot component of DeGaPe 35). The SED represents only a mean spectrum of DeGaPe 35 as the individual points were obtained over a relatively long period of time.

[TiO] index-spectral type relations using the spectra from the library of Fluks et al. (1994) down-sampled to the resolution of our spectra ($R \sim 550$). The spectral type corresponding to the obtained TiO indices is M3-5 in this case.

While this method is only very weakly dependent on the adopted reddening, it might provide a slightly earlier spectral type of the giant as the optical fluxes (especially at the shorter wavelengths) are typically influenced by the strong nebular radiation and/or the hot component. To have an independent estimate of the spectral type of the cool component of DeGaPe 35, we have directly compared the observed spectra with the ones from the library of Fluks et al. (1994). For calculation of $\chi^2$, only the red part of the spectra ($> 5\,500$ Å) was used, omitting the region of H$\alpha$ emission line. Both, reference and observed spectra were normalized to unity in the region $6\,300 - 6\,400$ Å. In addition, reference spectra were down-sampled to match the resolution of the observed spectra and reddened by various values in the interval $E(B-V) = 0.8 - 2.4$ mag (adopting the reddening law of Cardelli et al., 1989, and the total-to-selective absorption ratio $R = 3.1$). The upper limit corresponds to the total Galactic reddening in the direction of DeGaPe 35, $E(B-V) = 2.34$ mag, estimated from the dust map of Schlafly and Finkbeiner (2011)[4], while the lower limit is obtained from the comparison of observed and intrinsic infrared color indices of M-type giants (Jian et al., 2017).

The best fit was obtained for an M5 giant (Fig. 4) and the reddening value of $E(B-V) = 1.7$ mag. The extinction is similar to the value one would get for the intrinsic J-K colors of M5 giant (reddening $E(B-V) = 1.44$ mag calculated from $(J-K)_0 = 1.36$ and $(J-K) = 2.09$). The effective temperatures corresponding to the obtained spectral types are $3\,476$ K and $3\,367$ K for M4 and M5 giants, respectively (van Belle et al., 1999).

The similar temperature is suggested by the fitting of multi-frequency SED (Fig. 5) with the BT-Settl grid of theoretical spectra (Allard, 2014) downloaded from the Theoretical spectra webserver at the SVO Theoretical Model Services[5]. In the fitting procedure (for more details see Merc et al., 2020), only 2MASS, MSX and WISE observations were used, as the radiation at shorter wavelengths is strongly contaminated by the nebular emission and the hot component (the optical/near-UV excess is visible in the figure). We should note that this procedure is only very weakly dependent on the $\log g$ value, therefore we fixed the value on $\log g = 0$, which is a typical value for symbiotic giants with late spectral types (e.g., Gałan et al., 2016, 2017).

Assuming the luminosity class III, we can estimate the radius and luminosity of the cool component in DeGaPe 35. For M4 – M5 III stars, one can obtain the radii of $92-114\,R_\odot$ using the calibrations of van Belle et al. (1999). Luminosity in the range of $1\,100-1\,500\,L_\odot$ is then calculated using the Stefan–Boltzmann law. We should note that the assumption of the luminosity class III might not be true for all symbiotic stars (e.g., Mikołajewska, 2007).

The range of absolute bolometric magnitudes corresponding to the obtained values of the luminosity is $M_{\rm bol} = (-3.20) - (-2.86)$ mag. Using the relations of bolometric correction $BC_K$ and effective temperature (Buzzoni et al., 2010), we calculated $M_K = (-6.24) - (-5.81)$ mag. The dereddened $K$ magnitude of DeGaPe 35 from the 2MASS catalog is 5.63 mag. From apparent and absolute $K$ magnitude, one can get the distance estimate of $1.9 - 2.4$ kpc. This value is lower than the distance estimated independently using the *Gaia* EDR3 data by Bailer-Jones et al. (2021). They obtained geometric distance in the range of $3.5 - 5.3$ kpc and photogeometric distance of $3.1 - 4.3$ kpc. This might suggest that the real luminosity of the cool component in DeGaPe 35 is higher than the luminosity calculated under the assumption of luminosity class III. On the other hand, the goodness-of-fit of astrometric solution presented in *Gaia* EDR3 has a value of 3.37, which is higher than the limiting value of 3. Such solutions are suggested to be considered as unreliable in the *Gaia* online documentation. It is worth noting that only single star model was adopted for the astrometric solution in *Gaia* DR3 and the binarity might influence the astrometric parameters in such a long-period system (see, e.g., Sion et al., 2019). We have adopted the compromise value of 3 kpc as a distance to DeGaPe 35 in the rest of this work. That translates to the cool component luminosity of $2\,970 - 3\,220\,L_\odot$ and radius $150 - 167\,R_\odot$. We should however note that this is only a very rough estimate of the parameters which is dependent on the adopted distance, reddening, and assumes that the spectral type of the giant was inferred correctly.

---

[4]Unfortunately, it is not possible to use the 3D dust map of Green et al. (2019) as it covers only the sky north of a declination of -30 deg.

[5]http://svo2.cab.inta-csic.es/theory/newov2/index.php

### 4.2. Hot component

The contribution of the symbiotic white dwarfs to the observed continuum in the optical range of SED is typically negligible. For that reason it is not possible to infer the temperature and the luminosity of the hot component in DeGaPe 35 directly from the data we have available. However, one can obtain the rough estimates of the parameters using indirect methods which use the fact that the emission lines in the optical spectra are directly influenced by the high energy photons of the hot component.

The lower limit of the hot component temperature is given by the presence of emission lines with the highest ionization potential, i.e., Raman-scattered O VI lines in our case. According to the method of Murset and Nussbaumer (1994) that translates to the temperature of $\sim 114\,000$ K. The upper limit can be obtained from the fluxes of H I, He I, and He II lines (see Iijima, 1981), under the assumption of a black-body spectrum of the hot component and case B recombination. We have used slightly modified approach presented by Leedjärv et al. (2016), see their equation 2. This method uses equivalent widths of He II $\lambda$ 4686 and H$\beta$ emission lines and does not require the absolute flux calibration of the spectra which is not possible for our data due to the lack of simultaneous photometric observations. We should note that this method is only weakly dependent on the reddening as the ratio of the equivalent widths of lines relatively close to each other in the spectrum is used.

For the spectrum obtained in June 2021, we have obtained the temperature of the hot component of $T_h \sim 1.71 \times 10^5$ K, while the same analysis for the spectrum acquired in 2022 resulted in the value $T_h \sim 1.38 \times 10^5$ K. The value should not be overestimated by more than $15 - 20\%$ (see Merc et al., 2017). In general, such a range of temperatures is often observed in the case of symbiotic white dwarfs (see Fig. 4 in Mikołajewska 2003). The slight decrease in the temperature of the hot component between the observations is also consistent with the observed decrease of the fluxes of other highly ionized emission lines.

There are also indirect methods for the estimate of the hot component luminosity that employ the fluxes of He II $\lambda$ 4686, He I $\lambda$ 5876, and H$\beta$ (Kenyon et al., 1991; Mikolajewska et al., 1997). We have decided not to provide luminosity estimates for DeGaPe 35 since they would be calculated with a very large uncertainty due to the dependence on the imprecise absolute flux calibration, adopted reddening, and the distance. Moreover, the S/N in the blue part of the spectra is not very high in our observations (the star is highly reddened). That would influence the measurements of the flux in the lines as well.

On the other hand, the presence of strong emission lines of He II, [Fe VII], or O VI requires the luminosity of the hot component to be at least a few $\times 10^3$ $L_\odot$. The overall appearance of DeGaPe 35 in optical is similar to other well-known shell-burning symbiotic systems. Assuming the mass of the hot component in DeGaPe 35 of $0.6\,M_\odot$ (the median mass of symbiotic white dwarfs; Mikołajewska, 2007; Mikołajewska, 2010), the minimum luminosity of the shell-burning white dwarf would be about $2\,300 - 2\,400\,L_\odot$ (Nomoto et al., 2007; Wolf et al., 2013).

## 5. Conclusions

We have presented the analysis of low-resolution spectroscopic observations, supplemented by the photometry from *Gaia* DR3 and multi-frequency SED of DeGaPe 35. This object was previously classified as an emission-line star in the scope of an amateur survey searching for new planetary nebula candidates. Our results confirm that this source is a symbiotic star whose optical spectrum shows prominent emission lines, including highly ionized [Fe VII] and O VI lines. The cool component of this symbiotic binary is an M4-5 giant with $T_{\rm eff} \sim 3\,380 - 3\,470$ K and luminosity $\sim 3 \times 10^3\,L_\odot$ (for the adopted distance of 3 kpc). The infrared data of DeGaPe 35 allowed us to classify it as an S-type symbiotic star. The companion is most probably a shell-burning white dwarf. The photometric observations of the *Gaia* satellite, published recently in the *Gaia* DR3 suggested the variability with the period of about $700 - 800$ days, that we tentatively attributed to the orbital motion of the binary.

## Declaration of competing interests

The authors declare that they have no known competing financial interests or personal relationships that could have appeared to influence the work reported in this paper.

## Acknowledgments

We are thankful to an anonymous referee for the comments and suggestions improving the manuscript. This research was supported by the *Charles University*, project GA UK No. 890120, the internal grant VVGS-PF-2021-1746 of the *Faculty of Science, P. J. Šafárik University in Košice*, and the *Slovak Research and Development Agency* under contract No. APVV-20-0148.

# CRediT authorship contribution statement

**Thomas Petit:** Discovery of the object, Spectroscopic observation, Writing - original draft. **Jaroslav Merc:** Methodology, Analysis of the data, Writing - original draft, review & editing. **Rudolf Gális:** Writing - review & editing. **Stéphane Charbonnel:** Resources. **Thierry Demange:** Resources. **Richard Galli:** Resources. **Olivier Garde:** Resources. **Pascal Le Dû:** Resources. **Lionel Mulato:** Resources.